# New effects in absorption of ultrasound in intermediate state of high pure type I superconductor

A. G. Shepelev, O. P. Ledenyov and G. D. Filimonov

*National Scientific Centre Kharkov Institute of Physics and Technology,
Academicheskaya 1, Kharkov 61108, Ukraine.*

The research on the longitudinal ultrasound absorption in an intermediate state of the high pure *Ga* single crystal at the frequency of *50 – 190 MHz* at the temperatures of *0.37°K– 0.75°K*, using the impulse method, is conducted. The new effects in the absorption of ultrasound in an intermediate state of the high pure *type I* superconductor are discovered. The big oscillations in the dependence of the ultrasound absorption on the magnetic field *Γ(H)* in an intermediate state of the high pure *Ga* single crystal are experimentally observed at the magnitudes of magnetic field below the critical magnetic field, $H \lesssim H_c$. The maximum of the monotonic absorption of longitudinal ultrasound in an intermediate state of high pure *type I* superconductor is also obtained. The possible original theoretical mechanisms to explain the nature of big oscillations in the dependence *Γ(H)* in an intermediate state of the high pure *type I* superconductor are proposed.

## Introduction

The theoretically predicted phenomena [1] of the oscillatory absorption of ultrasound (*US*) in an intermediate state (*IS*) of the high pure *type I* superconductor [2], appearing at the change of the thickness of normal phase layers *a* as a result of the change of the magnitude of magnetic field *H*, can not be detected without overcoming the difficulties, connected with the creation of the fine *IS* periodic structure in the interior of the high pure *type I* superconductor, satisfying a number of strong inequalities between the electron mean free path *l*, thickness of normal phase layers *a*, and extreme diameter of electron orbit in magnetic field $D_{ext}$[1], namely

$$l \gg D_{ext} \gg a.$$

The criterion for satisfying the most stringent condition $l \gg D_{ext}$ (the magnetic field **H** has a defining influence on the dynamics of electrons in the normal phase layers) is the presence of natural magneto-acoustic oscillations in a normal state of the high pure *type I* superconductor at the temperature above the critical temperature $T > T_c$ and the magnetic field below the critical magnetic field $H < H_c$. The only superconductor in which one can certainly hope to observe the predicted phenomenon is the high pure *Gallium* (*Ga*) single crystal, in which the magneto-acoustic oscillations were observed in the normal state in the magnetic fields, starting from several *Oersted* [4].

## Discussion on measurements results

1. We present some experimental results, obtained at the research on the longitudinal ultrasound absorption in an intermediate state of the high pure *Ga* single crystal at the frequency of *50 – 190 MHz*, using the impulse method[2] [5]. The measurements were conducted in a cryostat, using the absorption pumping-off of the vapor of ~*25 cm³* of the liquid *Helium* ($^3He$) at the temperatures down to *0.32°K* [6]. The object of detailed research was a cylindrical sample with the diameter of *7 mm* and the length of *21 mm*, which was cut by the electric-spark method from the extremely high pure *Ga* single crystal, synthesized by the *Giredmet Experimental Plant*. The axis of the cylinder coincided with the crystallographic axis ***b*** of the high pure *Ga* single crystal and with the ultrasound propagation direction of ***k***[3]. The measurements of ultrasound absorption dependence *Γ(H)* were conducted by the relative method, comparing the propagated signal impulse with the standard impulse [5] in the regime of automatic plotting on the *PDS-021 x-y* recorder. The plates of *X*-cut quartz with the thickness of *0.3 mm* and the diameter of *4.5 mm* were used as the converters of the high frequency signal into the ultrasound (and vice-versa); it is necessary to note that the monochromatic beam of longitudinal ultrasound propagated along the central region of sample without the interaction with the side surfaces of a sample[4]. The intermediate state in the high pure *Ga* single crystal was created by the homogeneous transverse magnetic field *H*, using a pair of the *Helmholtz* coils ($k \perp H$). The vector *H* could be rotated in the plane of the axes *a* and *c* of the high pure *Ga* single crystal, which is perpendicular to the axis of a sample, at a rate of *1 rev/min*. The stabilization of the magnetic field *H* (and, when necessary, the change of the magnitude of magnetic field *H* in accordance with the linear law) was achieved by the magnetic field - stabilization and magnetic field - sweep schemes [7] with the stabilization accuracy better than $3\times10^{-4}$. During the course of the measurements, the sample was in direct contact with the liquid $^3He$, and the earth's magnetic field was compensated by the means of the two pairs of the *Helmholtz* coils.

2. In the experiment, we determined the dependence $\Gamma(H)$ in the high pure *Ga* single crystal at the constant temperature. Fig. 1 shows the dependence of the amplitude of transmitted ultrasonic signal on the magnetic field *A(H)* at the *Superconductor – Normal Metal* (*S-N*) and *Normal Metal – Superconductor* (*N-S*) transitions in an intermediate state of the high pure *Ga* single crystal at the temperature $T=0.39°K$, obtained with the *PDS-021* automatic recorder. The time to create each of the plots is around *45 min*. The big oscillations in the dependence of the ultrasound absorption on the magnetic field $\Gamma(H)$ in an intermediate state of the high pure *Ga* single crystal are observed at the magnetic field below the critical magnetic field $H \lesssim H_c$ in all the measurements, and they are independent of the frequency and amplitude of ultrasonic signal at the temperatures of *0.37 – 0.75 °K*. In view of the high purity of the *Ga* single crystal, the hysteresis phenomena are observed – the superheating (in the *S-N* transition) and the supercooling (in the *N-S* transition). However, the magnitude of the ultrasound absorption (even at the fastest rate of change of the magnetic field, *17 Oe/min*) is reversible with the accuracy of *0.2 dB/sample* in the *S-N* transition and back, presenting an evidence that there is no the "frozen-in" flux.

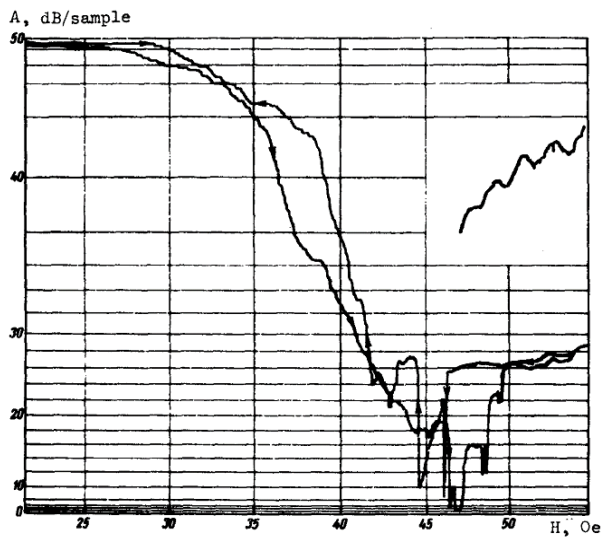

*Fig. 1. Dependence of amplitude of transmitted ultrasonic signal on magnetic field A(H) at S-N and N-S transitions in intermediate state of high pure Ga single crystal at temperature T=0.39°K, obtained with PDS-021 automatic recorder. Ultrasound frequency f=130 MHz, $k \perp H$, $k \parallel b$, $\angle H, c=22°$. Rate of change of H is 0.75 Oe/min. Insert shows plot of A(H) in normal state of high pure Ga single crystal at increased sensitivity of PDS-021 automatic recorder.*

To obtain the intermediate state structures, which are close to the equilibrium [2], we used the slow rotation of the magnetic field *H* between the points, separated by the following magnitude of magnetic field $\Delta H \simeq 0.003 H_c$. At a smooth variation of the magnetic field *H* with the rate of *0.7 Oe/min*, the vector *H* completed several revolutions around the axis of a sample at the rate of *1 rev/min*. It turned out that the very complicated oscillatory dependence $\Gamma(H)$ is obtained even after one revolution by magnetic field *H* in Fig. 2 (a). Since the magnitude of ultrasound absorption $\Gamma$ at each point after a change $\Delta H$ in the magnitude of magnetic field *H* does not depend on the time, it can be assumed that the created *IS* structure is close to the equilibrium; and it has, in accordance with the theory [2], the layered disc-like geometric shape. This is all the more correct, because the shape of curves $\Gamma(H)$ after the one and two revolutions by the magnetic field *H* are similar in Fig. 2., and the value of ultrasound absorption $\Gamma$ at each point after the two revolutions is practically independent of the number of revolutions[5]. The intermediate state regions with the small and big concentrations of the *N* phase with the probably present isolated filaments of the *N* and *S* phases and without the periodicity of the *IS* structure are characteristic. The fact that the magnetic field $H_c$ in the *N* phase layers doesn't change at the variation of magnetic field *H* results in a disappearance of the usual magneto-acoustic oscillations at the *N-S* transition. This is one more proof that the discovered effect can not be originated by the usual oscillatory phenomena, existing in the normal metal. The obtained results significantly differ from the data of previous ultrasonic investigations on the intermediate state of superconductors (see the review [9])[6]. Although the general picture of phenomena is complicated, the scale of the effect is striking, it make sense to note that the magnitude of certain oscillations exceeds the magnitude of total ultrasound absorption $\Gamma_0$ in the normal metal at the magnetic field equal to the critical magnetic field $H = H_c$! As shown by the computer modeling, at $k_a \sim 3\text{-}30$ and $a/D_{ext} \sim 0.1\text{-}0.9$, the scale of the real phenomenon is well described by the theory[7]. It is necessary to take into consideration that the experimental results may differ from the theoretical predictions, because the theory employed a model of superconductor with the very simple *Fermi* surface. The object of the research, the high pure *Ga* single crystal, has the complicated *Fermi* surface [12], and the phenomenon can apparently be the result of a superposition of several $D_{ext}$ (in the researched case: the two $D_{ext}$). In this connection, it is necessary to complete the theoretical calculations, considering the experimentally discovered interference picture of phenomenon[8]. The dimension *a*, in the volume of superconductor at the big concentrations of the *N* phase, estimated from the periodicity of the oscillations of the dependence $\Gamma(H)$, using the *Landau* non-branched model of an intermediate state of superconductor [2], amounts to $a \sim 10^{-2}$ cm.

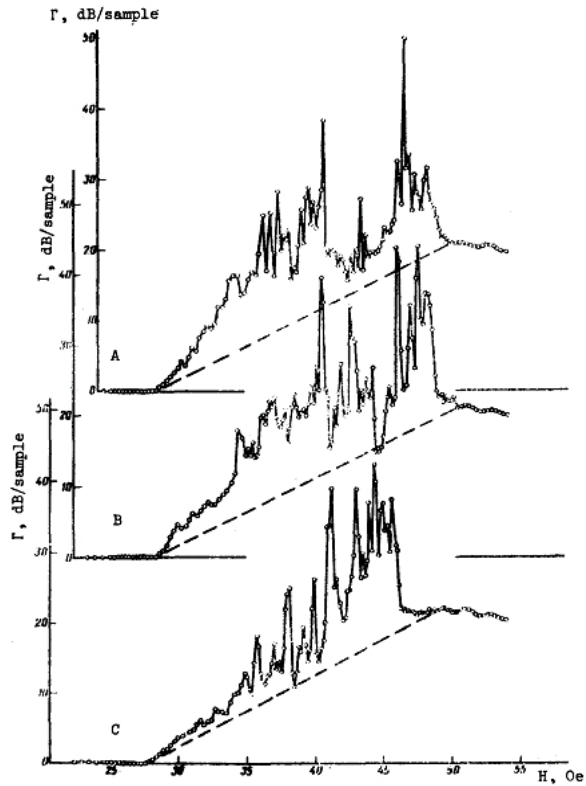

*Fig. 2. Dependence of ultrasound absorption on magnetic field $\Gamma(H)$ in intermediate state of high pure Ga single crystal:*
*A- after one revolution of magnetic field **H** between neighboring points at S-N transition;*
*B- after two revolutions of **H** at S-N transition;*
*C- after two revolutions of **H** at N-S transition.*
*Ultrasound frequency f=130 MHz,*
*$k \perp H, k \parallel b, \angle H, c = 22°$.*
*Time of each measurement exceeds 5 hours, temperature is close to 0.4 °K.*
*Dashed lines represent monotonic part of $\Gamma(H)$ at $k_a \gg 1$ in accordance with theory in [1].*

3. Besides the discovered giant oscillations in the dependence $\Gamma(H)$, we experimentally obtained a maximum of the monotonic absorption of longitudinal ultrasound in an intermediate state of high pure *type I* superconductor. The nature of this phenomenon is not clear yet, but the fact that the ultrasound absorption in an intermediate state of superconductor is in many times bigger in comparison with the ultrasound absorption in normal state of superconductor (and it is bigger than the ultrasound absorption, corresponding to the theory [1] at $k_a \gg 1$ (see Figs. 1 and 2)) represents a clear evidence about the strength of physical mechanism, causing this phenomenon. One of the probable mechanisms, namely the vibration by the inter-phase boundaries [13], can not seemingly give so big contribution to the magnitude of ultrasound absorption; here, it makes sense to note that the theoretical calculations [13] were performed at the low frequencies of ultrasonic signal ($10^6$ - $10^7$ Hz) and at $l \ll a$. It is possible that its contribution is considerable at the extreme conditions in our experiment ($k_a \gg 1$ and $l \gg a$).

## Conclusion

The research on the longitudinal ultrasound absorption in an intermediate state of the high pure *Ga* single crystal at the frequency of *50 – 190 MHz* at the temperatures of *0.37 °K–0.75 °K*, using the impulse method, is completed. The new effects in the absorption of ultrasound in an intermediate state of the high pure *type I* superconductor are discovered. The big oscillations in the dependence of the ultrasound absorption on the magnetic field $\Gamma(H)$ in an intermediate state of the high pure *Ga* single crystal are experimentally observed at the magnitudes of magnetic field below the critical magnetic field, $H \lesssim H_c$. The maximum of the monotonic absorption of longitudinal ultrasound in an intermediate state of high pure *type I* superconductor is also obtained. The possible original theoretical mechanisms to explain the nature of giant oscillations in the dependence $\Gamma(H)$ in an intermediate state of the high pure *type I* superconductor are proposed.

The detailed measurements to characterize the influence by the temperature, by the orientations of **k** and **H** relative to the crystallographic axes and relative to each other, and by the frequency of ultrasonic signal on the dependence of $\Gamma(H)$ in an intermediate state of the high pure *Ga* single crystal will be reported in the next complete research paper.

The authors are sincerely grateful to Alexander F. Andreev for the friendly thoughtful theoretical discussions during all the stages of innovative research, starting with the formulation of the new oscillatory phenomenon [1], to Yuri V. Sharvin for his physical and technical advices, to N. E. Alekseevskii, E. A. Kaner, B. G. Lazarev, V. L. Pokrovskii, and A. I. Shal'nikov for the numerous interesting discussions on the obtained research results.

This research paper was published in the *Letters to the Journal of Experimental and Theoretical Physics* (*JETP Letters*) in 1971 [14].

*E-mail: ledenyov@kipt.kharkov.ua

1) A. F. Andreev advised us that in the formulas in [1], the $R_{ext}$ should be considered to mean the extreme diameter $D_{ext}$ of electron orbit in the magnetic field **H**.

2) In our samples, at $k \perp H$, the magneto-acoustic oscillations in the normal state of the high pure *Ga* single crystal at the temperature $T > T_c$ and at the applied ultrasonic signal with the lowest frequency of *50 MHz* were observed, starting with the magnetic fields $H \simeq 3$ *Oe* (**k** is the wave vector of the ultrasonic wave).

3) The authors wish to express a deep gratitude to E. A. Levikov and L. V. Levikova for the precise orientation of researched samples with the help of the method of converging *X*-ray beams in a solid angle.

4) The authors are grateful to A. I. Shal'nikov and V. G. Bar'yakhtar for supplying the "No-nag" lubricant, which was used to make the acoustic contact between the sample and the quartz plates.

5) Walton [8], in an early study of the thermal conductivity, reached a similar conclusion, concerning the role of the rotation (at a rate of *0.1 rev/min*) in the formation of the equilibrium structures in an intermediate state of the *Tin*.

6) The observation of the non-monotonic dependence *Γ(H)* in an intermediate state of the *Tin* at the frequency of *19.5 MHz* was reported in [10]. However, the observed dependence *Γ(H)* had no the oscillations due to the fact that the condition $L >> D_{ext}$ was not satisfied, because, in the magnetic fields $H \leq H_c$ at the indicated frequency, the usual magneto-acoustic oscillations are not observed [11]: $D_{ext}=1.3 \times 10^{-1} - 3.2 \times 10^{-2}$ *cm* at the researched magnitudes of the critical magnetic field $H_c$ of the *Tin*, and the electron mean free path $l \approx 3 \times 10^{-2}$ *cm* [10]. In [10], the non-monotonicities appear as a result of complicated re-calculation, using the temperature dependences of several physical quantities, obtained for the various samples by different authors, and are within the limits of measurement and calculation accuracy.

7) The authors are grateful to L. G. Shepeleva for her help with the software development and calculations, using the "*Minsk-22*" computer.

8) We note that in an intermediate state, at any point, the sharp anisotropy of the coefficient of ultrasound absorption *Γ*, depending on the orientation of the magnetic field **H** in relation to the *Ga* axes (in the plane of the crystallographic axes ***a*** and ***c***) was found; - in some other directions the ultrasonic signal is masked by the noise (*Γ > 80 dB/sample* at the frequency of *130MHz*) – this is a result of real influence by the critical magnetic field $H_c$ on the electrons in the normal phase layers. Naturally, in the magnetic fields $H < H_c/2$, there is no dependence of the coefficient of ultrasound absorption *Γ* on the orientation of the magnetic field **H**.